\documentclass{aa}
\usepackage{graphicx}
\usepackage{txfonts}
\usepackage[]{hyperref}

\usepackage{dsfont} \usepackage{mathtools}

\usepackage[capitalise]{cleveref}
\Crefname{equation}{Eq.}{Eqs.}
\Crefname{figure}{Fig.}{Figs.}
\Crefname{tabular}{Tab.}{Tabs.}
\Crefname{section}{Sect.}{Sects.} \providecommand{\abs}[1]{\lvert#1\rvert}

\providecommand{\norm}[1]{\left\Vert#1\right\Vert}

 \newcommand{\td}{\,\mathrm{d}}  

\newcommand{\cronos}{\textsc{Cronos} }
\newcommand{\hllc}{\textsc{HLLC} }

\newcommand{\glorentz}{{u^0}}

\newcommand{\vU}{{\bf U}}
\newcommand{\vF}{{\bf F}}

\newcommand{\urad}{w_\text{rad}}
\newcommand{\umag}{w_B}
\newcommand{\Fsyn}{\text{F}}

\newcommand{\rfn}{g_n}
\newcommand{\rfe}{g_\epsilon}
\newcommand{\rfd}{g_\Omega} 
\begin{document}

\title{Relativistic fluid modelling of gamma-ray binaries}
\subtitle{I. The model}

\author{
   D. Huber \inst{1}
   \and
   R. Kissmann\inst{1}
   \and
   A. Reimer \inst{1}
   \and
   O. Reimer \inst{1}
}

\institute{
Institut f\"ur Astro- und Teilchenphysik \\
Leopold-Franzens-Universit\"at Innsbruck\\
6020 Innsbruck, Austria
\email{david.huber@uibk.ac.at}
}

\date{Received --; accepted --}

\abstract
{
Gamma-ray binaries are systems that radiate the dominant part of their non-thermal emission in the gamma-ray band.
In a wind-driven scenario, these binaries are thought to consist of a pulsar orbiting a massive star, accelerating particles in the shock arising in the wind collision.
}
{
We develop a comprehensive numerical model for the non-thermal emission of shock-accelerated particles including the dynamical effects of fluid instabilities and orbital motion.
We demonstrate the model on a generic binary system.
}
{
The model was built on a dedicated three-dimensional particle transport simulation for the accelerated particles that were dynamically coupled to a simultaneous relativistic hydrodynamic simulation of the wind interaction.
In a post-processing step, a leptonic emission model involving synchrotron and inverse-Compton emission was evaluated based on resulting particle distributions and fluid solutions, consistently accounting for relativistic boosting and $\gamma \gamma$-absorption in the stellar radiation field.
The model was implemented as an extension to the \cronos code.
}
{
In the generic binary, the wind interaction leads to the formation of an extended, asymmetric wind-collision region distorted by the effects of orbital motion, mixing, and turbulence. This gives rise to strong shocks terminating the pulsar wind and secondary shocks in the turbulent fluid flow.
With our approach it is possible for the first time to consistently account for the dynamical shock structure in particle transport processes, which yields a complex distribution of accelerated particles.
The predicted emission extends over a broad energy range, with significant orbital modulation in all bands.
}
{}


\maketitle

\section{Introduction} \label{sec:intro}
Gamma-ray binaries are composed of an early-type massive star in orbit with a compact object, either a neutron star or a black hole. They are characterised by a dominant radiative output in the gamma-ray regime $>1\,$MeV \citep[see][for a review]{Dubus2013A&ARv..21...64D, Paredes2019arXiv190103624P}.
They exhibit broadband non-thermal emission ranging from radio through low-energy (LE, 1-100 MeV) up to high-energy (HE, >100 MeV) and very high energy (VHE, >100 GeV) gamma-rays.
For most systems, observations show flux variations with the orbital phase.
At the time of writing, nine gamma-ray binaries emitting in the HE regime are confirmed:
1FGL J1018.6-5856 \citep{Fermi2012Sci...335..189F, Hess2015A&A...577A.131H},
4FGL J1405-6119 \citep{Corbet2019ApJ...884...93C},
HESS J0632+057 \citep{Aharonian2007A&A...469L...1A},
HESS J1832-093 \citep{HESS2015MNRAS.446.1163H, MartiDevesa2020A&A...637A..23M},
LMC P3 \citep{Corbet2016ApJ...829..105C, HESS2018A&A...610L..17H},
LS 5039 \citep{Paredes2000Sci...288.2340P, Aharonian2005Sci...309..746A},
LS I +61 303 \citep{Kniffen1997ApJ...486..126K, Albert2006Sci...312.1771A},
PSR B1259-63 \citep{Aharonian2005A&A...442....1A},
and PSR J2032+4127 \citep{Abeysekara2018ApJ...867L..19A}.
\\
In the literature, two possible mechanisms are proposed to explain the non-thermal emission \citep[see e.g.][]{Mirabel2006Sci...312.1759M, Romero2007A&A...474...15R}:
A jet-related emission scenario, where the compact object accretes matter from its stellar companion, releasing the accretion energy in the form of relativistic jets with high-energy particles \citep[see e.g.][]{Bosch-Ramon2009IJMPD..18..347B};
or a wind-driven scenario, in which the compact object is commonly assumed to be a pulsar, whose rotational energy is dissipated as a relativistic pair plasma interacting with the wind of the early-type companion star \citep[see][]{Maraschi1981MNRAS.194P...1M, Dubus2006A&A...451....9D}.
In close-orbit binaries, this interaction leads to the formation of an extended wind-collision region (WCR) that terminates both the pulsar and stellar wind by shocks \citep[see e.g.][]{Bogovalov2008MNRAS.387...63B, Bosch-Ramon2012A&A...544A..59B}.
At these sites, particles can be accelerated to ultra-relativistic energies through diffusive shock acceleration and other processes \citep{Sironi2011ApJ...741...39S}, which then produce the observed radiation.
For many systems, a wind-driven interpretation seems favoured.
This has been investigated using a variety of different approaches ranging from simple few-zone models to more complex numerical models including different emission mechanisms \citep[see e.g.][]{
  Khangulyan2007MNRAS.380..320K,
  Bosch-Ramon2008A&A...482..397B,
  Cerutti2010A&A...519A..81C,
  Bednarek2011MNRAS.418L..49B,
  Zabalza2013A&A...551A..17Z,
  Takata2014ApJ...790...18T,
  Dubus2015A&A...581A..27D,
  Barkov2018MNRAS.479.1320B,
  Molina2020arXiv200700543M}.
However, while the explored approaches succeed at modelling some observed features, they fail at reproducing others.
In this work, we also focus on a wind-driven scenario and aim to alleviate some of the encountered issues.
\\
In analogy to gamma-ray binaries, extended WCRs can also be formed in colliding-wind binaries (CWBs).
These systems are very similar in terms of wind interaction, but they harbour a second early-type star instead of the neutron star.
CWBs can therefore be viewed as a non-relativistic analogue of gamma-ray binaries in the wind-driven scenario.
The HE emission of such systems has been successfully described by \citet{Reitberger2017ApJ...847...40R} employing a dynamically coupled treatment of the particle transport and the fluid.
The corresponding numerical model was implemented as an extension to the \cronos code \citep[see][]{Kissmann2018ApJS..236...53K} solving a Fokker-Planck type particle transport equation for protons and electrons simultaneously to the equations of magneto-hydrodynamics.
\\
In this work, we extend this approach to gamma-ray binaries by translating the previous simulations into a relativistic framework, but we 
neglect the possible effect of the pulsar and stellar magnetic field on the fluid flow at this stage \citep{Dubus2015A&A...581A..27D,Bogovalov2019MNRAS.490.3601B}.
This allows a production of emitting particle distributions that is consistent with the dynamics of the wind interaction; this has not been attempted in the case of gamma-ray binaries before.
In turn, this enables the prediction of the non-thermal emission from gamma-ray binaries in a self-consistent framework.
\\
This is the first in a series of papers.
With this first paper, we establish the governing equations, describe their numerical treatment, and demonstrate the approach on a generic binary system.
In a second paper \citep{Huber2020-2}, the developed model is specifically applied to the LS 5039 system, and we compare the predicted emission to observations.
\\
The paper is structured as follows:
In \cref{sec:fluid model} and \cref{sec: particle transport} we introduce the governing equations for the coupled fluid and particle transport simulation, respectively, together with their numerical treatment and their implementation within \cronos.
In \cref{sec: gamma ray computation} we present the emission model involving synchrotron and inverse-Compton emission with additional modulations by relativistic boosting and $\gamma \gamma$-absorption.
In \cref{sec: generic binary} we demonstrate the general properties of our model on a generic binary system and present the resulting fluid structure, particle distributions, and emission spectra and light curves.
A concluding summary and outlook are given in \cref{sec: summary}. 
\section{Numerical model}\label{sec:numerical model}
The numerical model can be conceptually split into three distinct parts: 
the fluid model formulated in special relativity treating the interaction of the winds (see \cref{sec:fluid model}),  
the particle transport treating the evolution of the particle distributions embedded in the fluid (see \cref{sec: particle transport}), 
and the emission model predicting the radiation from the system (see \cref{sec: gamma ray computation}).
The first two are treated simultaneously in the simulation, whereas the radiation is computed in a post-processing step using previously obtained particle and fluid solutions.
\subsection{Fluid model}\label{sec:fluid model}
Owing to the relativistic pulsar wind, a classical treatment does no longer suffice and a special relativistic hydrodynamic framework has to be employed to accurately simulate the stellar and pulsar wind interaction.
In this section be briefly summarise the necessary ingredients that are treated in an extension of the \cronos code \citep{Kissmann2018ApJS..236...53K}, which will be described in more detail in a future paper.
The governing equations can be cast into a set of hyperbolic conservation equations that in Cartesian coordinates take the form
\begin{equation}
  \partial_t \vU + \partial_i \vF^i \left( \vU \right) = {\bf S},
  \label{eq: srhd conservation equation}
\end{equation}
where $\vU$ denotes the conserved quantities, $\vF^i$ its respective flux, and ${\bf S}$ an additional source term.
The first two can be expressed through primitive fluid variables such as the mass density $\rho$, the pressure $p$, the four-velocity $u^\mu$ , and specific enthalpy $h$ as
\begin{equation}
  \vU = \begin{pmatrix}
    D \\
    E \\
    m^j
  \end{pmatrix} =
  \begin{pmatrix}
    \rho \glorentz    \\
    D h \glorentz - p \\
    D h u^j
  \end{pmatrix}
  \label{eq: srhd conserved}
\end{equation}
and
\begin{equation}
  \vF^i(\vU) = \begin{pmatrix}
    D \,  u^i / u^0 \\
    m^i      \\
    m^j \,  u^i / u^0 + p \, \delta^{ij}
  \end{pmatrix} \label{eq: srhd fluxes}
.\end{equation}
In this paper we use the normalisation $c=1$ and notate the three spatial components of the fluid four-velocity vector as $\mathbf{u} = u^i$ and its Lorentz factor as $\glorentz$.
The presented system of five equations in six unknowns is closed by an additional equation of state.
For now we employ an ideal equation of state,
\begin{equation}
  h(\rho, p) = 1 + \frac{\Gamma}{\Gamma - 1} \frac{p}{\rho},
\end{equation}
where $\Gamma = 4/3$ is the constant adiabatic exponent of the fluid, as this was found to have only a negligible effect on the resulting fluid structure with respect to more sophisticate ones \citep[][]{Bosch-Ramon2015A&A...577A..89B}.
\\
In our simulation, we employed the \hllc solver \citep{Mignone2005MNRAS.364..126M} alongside a second-order sub-grid reconstruction using a \textsc{minmod} slope limiter.
To further increase the stability and robustness of our simulation, we directly evolved $\tau = E - D$, which replaces the energy equation, and solved an additional conservation equation for the specific entropy $s = p / \rho^\Gamma$ used in smooth flows,
\begin{equation}
  \partial_t (s D) + \partial_i (s D u^i / u^0) = 0.
\end{equation}

\subsubsection{Wind injection} \label{sec: wind injection}The stellar and pulsar wind were injected in the simulation by prescribing a solution in a spherical region with radius $r_\mathrm{inj}$ of the domain.
Inside each wind injection volume, we prescribed the solution as
\begin{equation}
  \rho \left( \mathbf{r} \right)  = \frac{\dot{M}}{4 \pi r^2 u} \, , \quad
  \mathbf{u}(\mathbf{r})      = u \, {\bf e}_r                \, , \quad
  p (\mathbf{r})                  = 10^{-6} \rho \left( \mathbf{r} \right),
\end{equation}
where $\mathbf{r}$ denotes the vector originating at the injection centre to a given point in space, $\dot{M}$ the mass injection rate, and $u$ the four-speed of the injected wind.
We kept the wind speed constant at its terminal velocity over the whole injection volume.
The density gradient was scaled such that the prescribed mass-loss rate was recovered for every spherical shell inside the injection region.
Because the effect of thermal pressure is neglected in our simulation, it was set to a low value scaled with the density gradient in order to keep the corresponding temperature constant.
\\
In simulations of gamma-ray binaries, it is common practice to prescribe the pulsar mass-loss rate through the ratio $\eta$ of the total momentum.
The pulsar mass-loss rate can hence be obtained by
\begin{equation}
  \dot{M}_p = \eta \frac{\dot{M}_s u_s}{u_p},
  \label{eq: pulsar mass loss rate}
\end{equation}
where quantities indicated with $p$ and $s$ belong to the pulsar and star, respectively.
\\
In practice, the injection was realised by resetting the states of the respective cells after every time-step with those computed from the prescribed solution.
A cell was treated as part of the star or pulsar, respectively, when its centre was within a spherical volume of given radius $r_\mathrm{inj}$ around the star or pulsar location.

\subsubsection{Orbital motion}\label{sec: orbital motion}
To account for the orbital motion, the wind injection volumes were relocated after every step of the simulation to their respective position on the Keplerian orbit.
In standard approaches, the system is treated in the laboratory frame \citep[see e.g.][]{Bosch-Ramon2012A&A...544A..59B}.
In our approach, we treat the binary in a frame corotating with the average angular velocity of the system as described in \cref{app: corotating frame}.
This has the advantage that the vector connecting the two companions remains fixed for circular orbits instead of performing full rotations.
For mildly eccentric orbits, this translates into an oscillation in a narrow angular range.
Consequently, the wind injection volumes have to be relocated by a smaller amount than in the laboratory frame, which increases the stability of the simulation by avoiding unphysical conditions upon relocation.
\\
Furthermore, because the binary motion is restricted, the orientation of the WCR is also restricted. 
This allows saving computational effort by reducing the simulation volume behind the star as seen from the centre of mass because we are mostly interested in the cometary tail-like downstream of the WCR behind the pulsar. \subsection{Particle transport}\label{sec: particle transport}
The particle transport equation for cosmic rays has been presented in the relativistic case by \citet{Webb1989ApJ...340.1112W}.
It takes the following form:
\begin{multline}
  0 = {\nabla_\mu \left( u^\mu f'_0 {+ {q^\mu}} \right) }
  + \frac{1}{p^2} \frac{\partial}{\partial p'} \left[
  \vphantom{\frac{1}{p'^2}}\right.
  {\vphantom{\int} - \frac{p'^3}{3} \nabla_\mu u^\mu f'_0 }
  {\vphantom{\int} + \dot{p'}_\mathrm{loss} p'^2 f'_0 }\\
  {\vphantom{\int} - {\Gamma_\mathrm{ visc} p'^4 \tau \frac{\partial f'_0}{\partial p'}}}
  {\vphantom{\int}
  - {p'^2 D_{pp} \frac{\partial f'_0}{\partial p'}} }
  {\vphantom{\int} - {p' (p'^0)^2 \dot{u}_\mu q^\mu} }
  \left. \vphantom{\frac{1}{p^2}} \right]
  ,
\end{multline}
\label{eq: transport webb}
where $f'_0(x^\mu, p')$ is the isotropic part of the particle phase space density.
All primed quantities are given in the fluid frame.
The first term describes passive advection with the fluid bulk motion $u^\mu$ and spatial diffusion with the heat flux $q^\mu$.
The other terms in square brackets describe the transport in momentum space.
From left to right, they correspond to adiabatic losses or gains, generic momentum losses (e.g. from radiative processes), viscous shear acceleration, the second-order Fermi process, and losses due to non-inertial frame changes.
Following \citet{Vaidya2018ApJ...865..144V}, we neglected (for the purpose of this paper) spatial diffusion ($q^\mu  = 0$), viscous shear-acceleration ($\Gamma_\mathrm{visc} = 0$), and momentum diffusion ($D_{pp}=0$).
Because we assume the ultra-relativistic limit, the particle Lorentz factor can be readily obtained as $\gamma \approx p' / m,$ leading us to the define $\mathcal{N}'(x^\mu, \gamma') = 4 \pi {p'}^2 f'_0 m$, with $\mathcal{N}' \td V' \td \gamma'$ corresponding to the number of particles within a volume $\td V'$ in the range $[\gamma',\gamma'+ \td \gamma']$.
According to our assumptions and substitutions, we obtain the simplified transport equation
\begin{equation}
  {\nabla_\mu \left( u^\mu \mathcal{N}' \right)}
  + \frac{\partial}{\partial \gamma'} \biggl[
    \underbrace{
      \left(
      - \frac{\gamma'}{3} \nabla_\mu u^\mu
      + \dot{\gamma'}_\mathrm{rad}
      \right)
    }_{= \langle \frac{\td \gamma'}{\td t'} \rangle}
    \mathcal{N}'
    \biggr]
  = 0, \label{eq: transport final}
\end{equation}
with $\langle \frac{\td \gamma'}{\td t'} \rangle $ being the total energy loss rate in the fluid frame.
We now consider electrons and positrons in the particle transport because accelerated nuclei are thought to not contribute efficiently to the overall emission \citep[e.g.][]{Bosch-Ramon2009IJMPD..18..347B}.

\subsubsection{Implementation}
Following \citet{Reitberger2014ApJ...782...96R}, we implemented the transport equation \cref{eq: transport final} within \cronos by employing a splitting scheme that separates the problem into a spatial and an energy part,
\begin{align}
  \partial_t \bigl( u^0 \mathcal{N}' \bigr) + \nabla_i \bigl( u^i \mathcal{N}' \bigr)         & = 0 \label{eq: spatial transport}    \\
  \partial_t \bigl( u^0 \mathcal{N}' \bigr)
  + \partial_{\gamma'} \left( \langle \frac{\td \gamma'}{\td t'} \rangle \mathcal{N}' \right) & = 0. \label{eq: energy transport pre}
\end{align}
Both parts are treated independently and sequentially one after the other.
The particle density was discretised in space in the same manner as the fluid variables, while its spectrum was subdivided into spectral bins $[\gamma'_{l-1/2}, \gamma'_{l+1/2}]$.
In our implementation, we allow the user to specify the limits of the simulated spectral range, that is, the first and the last edge $\gamma'_{-1/2}, \gamma'_{N_\gamma -1/2}$, and the number of spectral bins $N_\gamma$.
The range was then subdivided into logarithmic bins with
\begin{equation}
  \gamma'_{l-1/2} = \gamma'_{-1/2} \left( \frac{\gamma'_{N_\gamma-1/2}}{\gamma'_{-1/2}} \right)^{l/N_\gamma}
  , \; \mathrm{for} \, l \in \lbrace 0, \dots, N_\gamma \rbrace \, .
  \label{eq: energy grid}
\end{equation}
The cell centre $\gamma'_l$ is then obtained as the arithmetic mean of its limits.
\\
Averaging \cref{eq: spatial transport} over the $l$-th energy bin yields the finite-volume version of the spatial particle transport,
\begin{equation}
  {\nabla_\mu \left( u^\mu \mathcal{N}'_l \right)}
  = 0
\end{equation}\label{eq: spatial transport fv}with the cell averages $\mathcal{N}'_l = \Delta {\gamma'}_l^{-1} \int_{\gamma'_{l-1/2}}^{\gamma'_{l+1/2}} \mathcal{N}' \td \gamma'$ and $\Delta \gamma'_l = \gamma'_{l+1/2} - \gamma'_{l-1/2}$.
The spatial particle distribution corresponding to a given spectral bin can therefore be represented by an additional three-dimensional scalar field $\mathcal{N}'_l$.
The spatial transport, amounting to a mere spatial convection, is solved directly by the hydrodynamic solver similar to the mass continuity equation (the first component in \cref{eq: srhd conservation equation}).
\\
Averaging \cref{eq: energy transport pre} over a spatial cell with index $i,j,k$, and treating the fluid variables as constant in time yields
\begin{equation}
  \partial_t \mathcal{N}'_{i,j,k}
  + \partial_{\gamma'} \left( \frac{1}{u^0} \langle \frac{\td \gamma'}{\td t'} \rangle \mathcal{N}'_{i,j,k} \right) = 0, \label{eq: energy transport}
\end{equation}
where $\mathcal{N}'_{i,j,k}$ is the spatial cell average of the differential particle number density at a certain energy.
The spectral evolution of the particle distribution can therefore be viewed as an advection in energy with energy-dependent velocity $\dot{\gamma}' = \langle \frac{\td \gamma'}{\td t'} \rangle / u^0$.
\\
The evolution in energy was treated independently for each spatial cell by a semi-Lagrangian solver in analogy to the one presented by \citet{Zerroukat2006IJNMF..51.1297Z}, which we detail in the following.
For better readability, we omit in the following the prime and the indices on the particle number density $\mathcal{N} \equiv \mathcal{N}'_{i,j,k}$ and the particle Lorentz factor $\gamma \equiv \gamma'$.
We can show that the absolute number of particles in the energy range $[\gamma_-(t), \gamma_+(t)]$ remains constant
if the limits $\gamma_\pm(t)$ satisfy the characteristic equation
\begin{equation}
  \frac{\td}{\td t} \gamma = \dot{\gamma} (\gamma)
  ,
  \label{eq: characteristic equation}
\end{equation}
which can be exploited to construct a conservative and explicit update scheme for the particle number densities.
\\
We obtained the particle number density $\mathcal{N}^{n+1}_l$ at the next timestep $t^{n+1}$ by first considering where the particles inside the bin were advected from, that is, in which $\gamma$-range they were at time step $t^{n}$.
This was done by performing a backward integration of the bin edges $\gamma_{l\pm1/2}$ following \cref{eq: characteristic equation} to identify the corresponding values $\gamma^n_{l\pm1/2}$ at time $t^{n}$ (denoting propagated edges with a superscript).
\cref{eq: characteristic equation} was solved numerically using a second-order explicit Runge-Kutta method, for example Ralston's method, with the time step provided by the RHD solver.
\\
After the interval edges $\gamma^n_{l\pm1/2}$ were identified, the updated number density was obtained using particle conservation
\begin{equation}
  \Delta \gamma_l \, \mathcal{N}_l^{n+1}
= \int_{\gamma^n_{l-1/2}}^{\gamma^n_{l+1/2}} \mathcal{N}^{n} \td \gamma
  = I^n \left[ \gamma_{l+1/2}^{n} \right] - I^n \left[ \gamma_{l-1/2}^{n} \right],
  \label{eq: updated particle count}
\end{equation}
where $I^n\left[\gamma\right]$ denotes the cumulative particle count up to a value $\gamma$, for the spectrum at time $t^n$ (see also \cref{eq: cumulative particle count}).
To approximate the integral, we considered a second-order, piece-wise linear reconstruction in energy in each cell, in analogy to the spatial reconstruction in the hydrodynamics part,
\begin{equation}
  \mathcal{N}^n(\gamma) = \mathcal{N}^n_l + \delta^n_l (\gamma - \gamma_l)
  ,
\end{equation}
with $\gamma_{l-1/2} < \gamma < \gamma_{l+1/2}$.
Here, $\delta^n_l$ denotes the slope of the reconstruction obtained after applying the van Leer limiter \citep{VanLeer1977JCoPh..23..276V},
\begin{equation}
  \delta_\mathrm{van Leer} = \frac{\max \left(\delta^+ \delta^-, 0\right)}{\delta^C}
  \label{eq: slope limiter}
,\end{equation}
to the right-, left-, and two-sided finite differences, denoted by $\delta^+, \delta^-$ , and $\delta^C$, respectively.
With this, the integral can be approximated directly as piecewise parabolas by
\begin{multline}
  I^n \left[ \gamma \right] =
  \int_{\gamma_{-1/2}}^{\gamma} \mathcal{N}^n(\tilde{\gamma}) \td \tilde{\gamma}
  \\
  \approx  I^n \left[ \gamma_{l-1/2} \right] + (\mathcal{N}^n_l - \delta^n_l \gamma_l) h_\gamma + \frac{\delta^n_l}{2} \, h_\gamma^2,
  \label{eq: cumulative particle count}
\end{multline}
with $\gamma_{l-1/2} < \gamma < \gamma_{l+1/2}$, $h_\gamma = \gamma -  \gamma_{l-1/2}$ , and the exact integrals at the cell edges,
\begin{equation}
  I^n \left[ \gamma_{l-1/2} \right] = \int_{\gamma_{-1/2}}^{\gamma_{l-1/2}} \mathcal{N}^{n} \td \gamma = \sum_{j = 0}^{l-1} \Delta \gamma_j \, \mathcal{N}_j^{n},
\end{equation}
where $\gamma_{-1/2}$ is the lower edge of the first spectral bin $l=0$.
\\
On either end of the energy grid, we employed outflowing boundary conditions, ensuring that particles cannot be gained but only lost through the boundaries.
This was realised by setting the particle number density to zero outside the energy domain, which amounts to a constant extrapolation of $I^n$ in the evaluation of the right-hand side of \cref{eq: updated particle count}.
To determine the slope of the first and last spectral bin according to \cref{eq: slope limiter}, we added a ghost-cell at either end of the spectrum (with $l=-1$ and $l=N_\gamma$, using \cref{eq: energy grid}), and set the cell values to zero.

\subsubsection{Magnetic field model}\label{sec: magnetic field model}
For the purpose of this work, we employed a simplified magnetic field model.
We assumed the magnetic field to be amplified by plasma instabilities to a fraction $\zeta_b$ of the available internal energy.
This fraction was kept constant throughout the simulation \citep[see also][]{Dubus2015A&A...581A..27D,Barkov2018MNRAS.479.1320B}.
The strength of the post-shock magnetic field $B'$ in the fluid frame can therefore be expressed as
\begin{equation}
  \frac{B'^2}{2 \mu_0} = \zeta_b \frac{p}{\Gamma - 1}
.\end{equation}
We assumed no direction on the magnetic field.

\subsubsection{Radiative losses} \label{sec: radiative losses}
As energy-loss processes for the accelerated electron-positron pairs we considered synchrotron and inverse-Compton losses.
The energy-loss rates are given by
\begin{equation}
  \dot{\gamma}'_\mathrm{syn} = \frac{4}{3} \frac{\sigma_T}{m_e c} \umag' {\gamma'}^2
  \label{eq: synchrotron loss rate}
\end{equation}
\begin{equation}
  \dot{\gamma}'_\mathrm{ic} = \frac{4}{3} \frac{\sigma_T}{m_e c} \urad' {\gamma'}^2 F_\mathrm{KN}(\epsilon_0', \gamma')
  \label{eq: inverse compton loss rate}
,\end{equation}
with $\umag'$ and $\urad'$ being the magnetic and radiative energy densities in the fluid frame, respectively, and $F_\mathrm{KN}$ the Klein-Nishina factor.
\\
The magnetic field in the fluid frame can be readily obtained by the transformation
\begin{equation}
\mathbf{B}' = \frac{1}{\glorentz} \left[ \mathbf{B} + \frac{\mathbf{B} \cdot \mathbf{u}}{\glorentz + 1} \mathbf{u} \right]
  \label{eq: comoving mag field}
.\end{equation}
The corresponding magnetic field strength is therefore given by
\begin{equation}
B' = \frac{1}{\glorentz} \sqrt{ B^2 + \left(\mathbf{B} \cdot \mathbf{u}\right)^2}
  \label{eq: comoving mag strength}
,\end{equation}
yielding the magnetic energy density $\umag' = \frac{B'^2}{2 \mu_0}$.
For disordered magnetic fields, that is, when no information on the magnetic field orientation is available, we dropped the term containing $\mathbf{B} \cdot \mathbf{u}$ in \cref{eq: comoving mag strength} \citep[see][]{Dubus2015A&A...581A..27D}.
\\
For the inverse-Compton losses, we assumed the radiation field to be monochromatic and mono-directional.
The energy density of a beam of photons with direction $\mathbf{\Omega}_0$ is given in the fluid frame by Doppler boosting the laboratory energy density,
\begin{equation}
  \urad' = \mathcal{D}_{u \Omega_0}^{-2} \urad
,\end{equation}
with the Doppler factor $\mathcal{D}_{u \Omega_0} = \left( \glorentz - \mathbf{u} \cdot \mathbf{\Omega}_0 \right)^{-1}$.
The radiation field relevant for the inverse-Compton losses emerges from the stellar binary companion,
\begin{equation}
  \urad = \frac{L_\mathrm{star}}{4 \pi r_\mathrm{star}^2 c},
\end{equation}
where $r_\star$ is the distance to the star and $L_\star$ its luminosity.
We employed the Klein-Nishina factor following \citet{Moderski2005MNRAS.363..954M},
\begin{multline}
  F_\mathrm{KN}(\tilde{b}') = \frac{9}{{\tilde{b'}}^3} \left[
  \left( \frac{1}{2} \tilde{b'} + 6 + \frac{6}{\tilde{b'}} \right) \log(1 + \tilde{b'})
  + 2 \mathrm{Li}_2(-{\tilde{b'}})
  \right. \\
  \left. - \left( \frac{11}{12} {\tilde{b'}}^3 + 6 {\tilde{b'}}^2 + 9 {\tilde{b'}} + 4 \right) \cdot \left( 1 + {\tilde{b'}} \right)^{-2}
  - 2
  \right]
  ,
\end{multline}
assuming an isotropic seed photon and electron distribution for simplicity, with $\tilde{b}' = 4 \epsilon'_0 \gamma' \, (m_e c^2)^{-1}$ and $\epsilon'_0 = \mathcal{D}_{u \Omega_0}^{-1} \epsilon_0$ denoting the Doppler-boosted photon energy of the stellar radiation field.

\subsubsection{Particle injection}\label{sec: particle injection}
Strong shocks are commonly regarded as sites for particle acceleration, where we expect a fraction of pairs from the pulsar wind to be accelerated to high energies.
However, because the converging fluid flow is the principal driver for several acceleration processes, we generally consider that strongly compressive flows participate in the acceleration.
In the simulation, we identified these sites using the criterion
\begin{equation}
  \nabla_\mu u^\mu < \delta_\mathrm{sh},
  \label{eq: injection criterion}
\end{equation}
with a typical value of $\delta_\mathrm{sh} = -10 \,$c$/$AU to ignore weakly converging flows (see also \cref{app: injection criterion} for a comparison to a common shock-identification criterion).
\\
In the identified cells we injected two populations of pairs: thermal pairs from the cold pulsar wind that are isotropised, following a Maxwellian distribution; and non-thermal accelerated pairs, following a power-law spectrum \citep[similar to][]{Dubus2015A&A...581A..27D}, which are parametrised as
\begin{equation}
  \mathcal{N}^\mathrm{MW} = K^\mathrm{MW} \gamma^2 \exp(-\gamma / \gamma_t)
\end{equation}
\begin{equation}
  \mathcal{N}^\mathrm{PL} = K^\mathrm{PL} \gamma^{-s} \quad \mathrm{for} \quad \gamma_\mathrm{min}^\mathrm{PL} < \gamma < \gamma_\mathrm{max}^\mathrm{PL}
.\end{equation}
The maximum Lorentz factor $\gamma_\mathrm{max}^\mathrm{PL}$ for the power law is determined by the balance between acceleration and radiative losses.
The acceleration timescale is given by a multiple of the gyration time of the particle $\tau_\mathrm{acc} = 2 \pi R_L / c \xi_\mathrm{acc}$ with the Larmor radius $R_L = \frac{\gamma m}{e B'}$ and the acceleration efficiency $\xi_\mathrm{acc}$.
We assumed synchrotron losses to be dominant for the particles at highest energies.
Together with the loss timescale $\tau_\mathrm{loss} = \gamma' / \dot{\gamma}'$ given by \cref{eq: synchrotron loss rate}, this yields the maximum particle Lorentz factor expected at the current location,
\begin{equation}
  \gamma_\mathrm{max}^\mathrm{PL} = \sqrt{
    \frac{3}{4 \pi}
    \frac{\mu_0 e c}{\xi_\mathrm{acc} \sigma_T B'} }
  \approx 4.65 \times 10^7 \sqrt{\frac{1\,\mathrm{G}}{\xi_\mathrm{acc} B'}}
  \label{eq: injection gamma max}
.\end{equation}
\\
Because the dominant particle acceleration mechanism in gamma-ray binaries could not be clearly identified, we took a generic phenomenological approach and treated both the spectral index $s$ and the acceleration efficiency $\xi_\mathrm{acc}$ as free parameters of the model \citep[similar to][]{Dubus2015A&A...581A..27D, Molina2020arXiv200700543M}.
\\
The average Lorentz factor of the Maxwellian $\gamma_t$, the lower cutoff of the power-law spectrum $\gamma_\mathrm{min}$, and both normalisations $K^{\mathrm{PL}\mathrm{MW}}$ were determined from fluid properties.
This was controlled by three free parameters: $\zeta_n^\mathrm{PL}$ the fraction of accelerated non-thermal pairs, $\zeta_e^\mathrm{PL}$ the fraction of internal energy converted to non-thermal pairs, and $\zeta_\rho$ to account for the pulsar wind speed, which might be too low in our simulation and result in an overestimation of the particle number density in the pulsar wind.
The respective fractions for the Maxwellian are then readily given by $\zeta_n^\mathrm{MW} = 1- \zeta_n^\mathrm{PL}$ and $\zeta_e^\mathrm{MW} = 1 - \zeta_e^\mathrm{PL}$.
In our simulation, we kept all injection parameters constant in time and space.
The injected spectra are therefore related to fluid quantities by
\begin{equation}
  \int_{\gamma_\mathrm{min}^\mathrm{PL,MW}}^{\gamma_\mathrm{max}^\mathrm{PL,MW}} \mathcal{N}^\mathrm{PL,MW}(\gamma) \, \td\gamma =
  \underbrace{\zeta^\mathrm{PL,MW}_n \, \zeta_\rho\, \psi_\mathrm{p} \, \rho / m_e}_{=: n^\mathrm{PL,MW} }
  \label{eq: injection particle number}
\end{equation}
\begin{equation}
  \int_{\gamma_\mathrm{min}^\mathrm{PL,MW}}^{\gamma_\mathrm{max}^\mathrm{PL,MW}} \gamma \, m_e c^2 \, \mathcal{N}^\mathrm{PL,MW}(\gamma)  \td\gamma =
  \underbrace{\zeta^\mathrm{PL,MW}_e \, \psi_\mathrm{p} \, e_\mathrm{thermal}}_{=: e^\mathrm{PL,MW}}
  \label{eq: injection energy}
,\end{equation}
where $0 < \psi_p < 1$ is a passive tracer indicating the fraction of pulsar wind material and the thermal energy density $e_\mathrm{thermal} = \frac{p}{\Gamma - 1}$ given by the fluid pressure.
\\
For the non-thermal particles, the above expressions yield an implicit equation for $\gamma_\mathrm{min}^\mathrm{PL}$, which was solved numerically by a standard root-finding algorithm.
The injection normalisation $K^\mathrm{PL}$ was then determined from \cref{eq: injection particle number}.
When the internal fluid energy was too low,
\begin{equation}
  \int_{\gamma_\mathrm{min}^\mathrm{PL}}^{\gamma_\mathrm{max}^\mathrm{PL}} \gamma \, m_e c^2 \, \mathcal{N}^\mathrm{PL}(\gamma)
  \, d\gamma
  > e^\mathrm{PL}
,\end{equation}
we did not inject particles in the given cell.
\\
For the Maxwellian, \cref{eq: injection particle number,eq: injection energy} can be inverted analytically in the limit $\gamma_t \gg 1$ using $\gamma_\mathrm{min}^\mathrm{MW} = 1$ and $\gamma_\mathrm{max}^\mathrm{MW} = \infty$, yielding
\begin{equation}
  \gamma_t = \frac{1}{3} \frac{e^\mathrm{MW}}{n^\mathrm{MW} \, m_e c^ 2 }, \; K^\mathrm{MW} = \frac{n^\mathrm{MW}}{2 \gamma_t^3}
.\end{equation} \subsection{Emission}\label{sec: gamma ray computation}
The particle distributions resulting from a joint hydrodynamic and particle-transport simulation (as detailed in \cref{sec:fluid model,sec: particle transport}) were used in a post-processing step to predict the expected emission.
In this section, we elaborate on the calculations of the emission and their implementation.
\\
The involved radiative processes are again synchrotron (\cref{sec: synchrotron emission}) and inverse-Compton emission (\cref{sec: inverse Compton emission}).
Furthermore, the produced fluxes were modified by $\gamma \gamma$-absorption on photons of the stellar radiation field depending on the location of the emitter (\cref{sec: absorption}).
For the sake of better readability, we omit the dependence on the location $\mathbf{x}$ when it is not imperative.
\\
We formulated the emission in the fluid frame, represented by primed quantities, and then related it to the laboratory frame with a Doppler boost.
The spectral emissivity $j_\epsilon$, that is, the emitted power per unit photon energy per unit solid angle per unit volume, for a given emitted photon energy $\epsilon$ and direction $\mathbf{\Omega}$ is hence given by
\begin{equation}
  j_\epsilon (\epsilon, \mathbf{\Omega}) = \mathcal{D}_{u \Omega}^2 j'_\epsilon (\epsilon', \mathbf{\Omega}')
,\end{equation}
where $\mathcal{D}_{u \Omega} = \left( \glorentz - \mathbf{\Omega} \cdot \mathbf{u} \right)^{-1}$ is the corresponding Doppler factor.
The respective energy and direction of the emitted photons are given in the fluid frame by
\begin{equation}
  \epsilon' = \mathcal{D}_{u \Omega}^{-1} \epsilon \, , \;
  \mathbf{\Omega}' = \mathcal{D}_{u \Omega} \left[ \mathbf{\Omega} + \left( \frac{\mathbf{\Omega} \cdot \mathbf{u}}{\glorentz + 1} - 1\right) \mathbf{u} \right]
.\end{equation}
Furthermore, it is convenient to express the spectral emissivity in terms of the emission produced by a single electron $\mathcal{P}'(\epsilon', \mathbf{\Omega}', \gamma')$, that is, the emitted power per unit photon energy per unit solid angle per electron under the assumption of an isotropic distribution of particles in the fluid frame.
The spectral emissivity can then be formulated as
\begin{equation}
  j'_\epsilon (\epsilon', \mathbf{\Omega}') = \int \mathcal{P}'(\epsilon', \mathbf{\Omega}', \gamma') \mathcal{N}' \td \gamma'
.\end{equation}

\subsubsection{Synchrotron emission}\label{sec: synchrotron emission}
The total power emitted by a single electron in a magnetic field $\mathbf{B}'$ can be expressed as
\begin{equation}
  {\mathcal{P}'}^\text{tot}_\text{syn}(\epsilon', \gamma') = \frac{\sqrt{3} e^3}{8 \pi^2 \varepsilon_0 m_e c \hbar} B'_\perp  \Fsyn\left(\frac{\epsilon'}{\epsilon_c'} \right)
  \label{eq: synchrotron total emission}
,\end{equation}
which is peaked around the critical photon energy,
\begin{equation}
  \epsilon_c' = \frac{3 e \hbar}{2 m_e} B'_\perp {\gamma'}^2 = 1.74 \times 10^{-8} \, \text{eV} \, \frac{B'_\perp}{1\,\text{G}} {\gamma'}^2
  \label{eq: synchrotron characteristic energy}
.\end{equation}
$B'_\perp = B' \sin(\alpha)$ denotes the magnetic field component perpendicular to the electron motion, which gyrates with a pitch angle $\alpha$ around the magnetic field lines.
For ultra-relativistic electrons, the emission is strongly beamed in their instantaneous direction of motion (within a cone of angle $1/\gamma'$).
Only particles whose pitch angle coincides with the angle between the magnetic field $\mathbf{B}'$ and the line of sight $\mathbf{\Omega}'$ therefore contribute to the emission, effectively yielding $B'_\perp = \abs{\mathbf{B}' \times \mathbf{\Omega}'}$.
Together with the assumption of an isotropic particle distribution, this yields
\begin{equation}
  \mathcal{P}'_\text{syn}(\epsilon', \mathbf{\Omega}', \gamma')
  = \frac{\sqrt{3} e^3}{32 \pi^3 \varepsilon_0 m_e c \hbar} B'_\perp \Fsyn\left(\frac{\epsilon'}{\epsilon_c'} \right)
  \label{eq: synchrotron single emissivity}
.\end{equation}
In many cases, we approximated the magnetic field to be disordered, that is, we have no information on the direction of the field lines.
In this case, we approximated the $B'_\perp$ in \cref{eq: synchrotron characteristic energy,eq: synchrotron single emissivity} by its pitch-angle average $\bar{B}'_\perp = \frac{\pi}{4} B'.$

\subsubsection{Inverse-Compton emission}\label{sec: inverse Compton emission}
Through the inverse-Compton process, photons of the stellar radiation field are scattered to gamma-ray energies by highly energetic particles.
Following the description of \citet{Moderski2005MNRAS.363..954M}, the power emitted through this process is given by
\begin{equation}
  \mathcal{P}'_\text{ic}(\epsilon', \mathbf{\Omega}', \gamma')
  = \epsilon' \frac{\partial \dot{N}'_\text{sc}}{\partial \epsilon' \, \partial \Omega'}
,\end{equation}
where $\frac{\partial \dot{N}'_\text{sc}}{\partial \epsilon' \, \partial \Omega'}$ is the scattering rate of photons on isotropically distributed electrons.
The scattering rate was presented by \citet{Aharonian1981Ap&SS..79..321A} in the case of $\epsilon' \gg \epsilon'_0$ and $\gamma' \gg 1$ taking the form
\begin{multline}
  \frac{\partial \dot{N}'_\text{sc}}{\partial \epsilon' \, \partial \Omega'}
  =  \frac{3}{16 \pi} c \sigma_T \, \frac{1}{{\gamma'}^2} \\
  \iint_{\epsilon'_{0,\text{min}}}^\infty
  f_\text{ic}(\epsilon', \epsilon'_0, \gamma', \mu')
  \frac{n_{\epsilon_0}'(\epsilon_0',  \boldsymbol{\Omega}_0')}{\epsilon_0'}
  \td \epsilon'_0 \td^2 {\Omega_0}' ,
  \label{eq: ic scattering rate}
\end{multline}
where $n_0(\epsilon_0, \mathbf{\Omega}_0)$ is the density of the seed photons with energy $\epsilon_0$ and direction $\boldsymbol{\Omega}_0$ and ${\epsilon_{0,\text{min}}'}$ the lowest seed photon energy that is still scattered to a given energy $\epsilon'$ and direction $\mathbf{\Omega}'$
\begin{equation}
  {\epsilon_{0,\text{min}}'} = \epsilon' \left[ 2 (1-\mu')(1-\frac{\epsilon'}{\gamma' m_e c^2}) \right]^{-1}
.\end{equation}
The scattering kernel $f_\text{ic}$ is given by
\begin{equation}
  f_\text{ic}(\epsilon', \epsilon'_0, \gamma', \mu') = 1 + \frac{{w'}^2}{2(1-{w'})} 
  - \frac{2 {w'}}{b'_\mu(1-w')} + \frac{2 {w'}^2}{{b'_\mu}^2(1-w')^2},
\end{equation}
with $b'_\mu = 2 (1 - \mu') \epsilon'_0 \gamma' (m_e c^2)^{-1}$, $w' = \epsilon' (\gamma' m_e c^2)^{-1}$ and the cosine of the scattering angle $\mu ' = \boldsymbol{\Omega}' \cdot \mathbf{\Omega}'_0$, which can be expressed in terms of laboratory quantities by
\begin{equation}
  \mu' = 1 + \mathcal{D}_{u \Omega_0} \mathcal{D}_{u \Omega} \left( \mu - 1 \right)
.\end{equation}
Together with the relativistic invariant $\frac{n_{\epsilon_0}'}{\epsilon_0'}
  \td \epsilon'_0 \td^2 {\Omega_0}' = \frac{n_{\epsilon_0}}{\epsilon_0}
  \td \epsilon_0 \td^2 {\Omega_0}$
\cref{eq: ic scattering rate} can be expressed directly in terms of the seed photon field in the laboratory frame,
\begin{multline}
  \frac{\partial \dot{N}'_\text{sc}}{\partial \epsilon' \, \partial \Omega'}
  = \frac{3}{16 \pi} c \sigma_T \, \frac{1}{{\gamma'}^2} \\
  \iint_{\epsilon_{0,\text{min}}}^\infty
  f(\epsilon', \epsilon'_0, \gamma', \mu')
  \frac{n_{\epsilon_0}(\epsilon_0,  \boldsymbol{\Omega_0})}{\epsilon_0}
  \td \epsilon_0 \td^2 {\Omega_0.}
\end{multline}
Applying the factorisation of the target photon field (\cref{sec: radiation fields}) yields
\begin{multline}
  \frac{\partial \dot{N}'_\text{sc}}{\partial \epsilon' \, \partial \Omega'}
  = \frac{3}{16 \pi} c \sigma_T \, \frac{1}{{\gamma'}^2}
  \rfn(\mathbf{x})
  \int
  \rfd(\mathbf{\Omega}_0)\\
  \int_{\epsilon_{0,\text{min}}}^\infty
  f(\epsilon', \epsilon'_0, \gamma', \mu')
  \frac{\rfe(\epsilon_0)}{\epsilon_0}
  \td \epsilon_0
  \td^2 {\Omega_0},
\end{multline}\label{eq: ic scattering rate final}
which can be further simplified depending on the employed radiation field model.

\subsubsection{$\gamma\gamma$-absorption}\label{sec: absorption}
As a result of the strong stellar photon field, parts of the VHE gamma-ray flux produced through inverse-Compton scattering are significantly attenuated by $\gamma \gamma$-pair creation.
This effect depends strongly on the geometry of the system and the location of the observer, yielding strongly modulated spectra with the binary orbit.
For a consistent model of gamma-ray binary emission, this attenuation has to be taken into account for every cell in our simulation.
\\
The absorption opacity $\tau_{\gamma\gamma}$ of photons with energy $\epsilon$ originating at $\mathbf{x}$ in direction $\mathbf{\Omega}$ can be expressed by a three-fold integration of the differential absorption opacity given by \citet{Gould1967PhRv..155.1404G},
\begin{equation}
  \td \tau_{\gamma\gamma} = \left( 1 - \mu \right) \sigma_{\gamma \gamma}(\beta) n_{\epsilon_0}(\mathbf{x}, \epsilon_0, \mathbf{\Omega}_0) \td l \td^2 \Omega_0 \td \epsilon_0
,\end{equation}
over the line of sight $l$, the solid angle $\Omega_0$ , and the seed photon energy $\epsilon_0$.
Here, $\mu = \mathbf{\Omega} \cdot \mathbf{\Omega}_0$ denotes the scattering angle cosine between a VHE photon with direction $\mathbf{\Omega}$ and a seed UV photon with direction $\mathbf{\Omega}_0$, $\beta^2 = 1 - \frac{m_e^2 c^4}{\epsilon \epsilon_0} \frac{2}{1 - \mu}$ the speed of the created $e^\pm$ pair normalised to $c$ and $\sigma_{\gamma \gamma} = \frac{3}{16} \sigma_T \, f_{\gamma\gamma}(\beta)$ the scattering cross-section with
\begin{equation}
  f_{\gamma\gamma}(\beta)  = (1-\beta^2) \left[ \left( 3 - \beta^4 \right) \log\left( \frac{1 + \beta}{1 - \beta}\right) - 2 \beta (2 - \beta^2) \right]
.\end{equation}
The lower bound of the energy integration is provided by the lowest seed photon energy $\epsilon_{0,\text{min}} = \frac{m_e^2 c^4}{\epsilon} \frac{2}{1-\mu}$ to create an $e^\pm$ pair under a given scattering angle.
The stellar seed photon density $n_{\epsilon_0}$ is defined in the same way as in \cref{sec: inverse Compton emission}.

For numerical purposes, it is more suitable to transform the occurring improper integrals into proper ones.
For the line-of-sight integration, we therefore used the transformation $l = d_0 \frac{\sin(\psi - \psi_0)}{\sin(\psi)}$ \citep[as detailed in][]{Dubus2006A&A...451....9D}.
The integration over target photon energy $\epsilon_0$ was transformed into an integral over the speed of the created pairs $\beta$.
Together with the radiation field factorisation (\cref{sec: radiation fields}), this yields
\begin{multline}
  \tau_{\gamma\gamma} =
  \frac{3}{16} \sigma_T
  \int_{\psi_0}^\pi \frac{\td l}{\td \psi} \rfn(\mathbf{x}')
  \int (1-\mu) \rfd(\mathbf{x}', \mathbf{\Omega}_0)\\
  \int_{0}^1
  \frac{\td \epsilon_0}{\td \beta}
  \rfe(\mathbf{x}', \epsilon_0) f_{\gamma \gamma}(\beta)
  \td \beta
  \td^2 \Omega_0
  \td \psi,
\end{multline}\label{eq: opacity final}
which can be further simplified depending on the stellar radiation field model.
\\
The integration over solid angle $\Omega_0$ is approximated by a nested quadrature rule with equidistant points in azimuth and a Gauss-Legendre quadrature in the polar angle \citep[see][]{Hesse2012}.
For the integrals over $\psi$ and $\beta,$ we employed standard Gauss-Legendre quadratures.
The necessary routines are provided by the \textsc{GSL}.

\section{Application to a generic binary}\label{sec: generic binary}
In the following section, we apply the presented numerical model to a generic binary system.
We focus on exploring the general properties of our model.
\\
The considered binary is composed of an O-type star in a circular ($e=0$) $1\,$d orbit with a pulsar at an orbital separation of $d=0.2\,$AU.
We adopted a stellar mass-loss rate of $\dot{M}_s = 2 \times 10^{-8} \,\text{M}_\odot\, \text{yr}^{-1}$ launched at a velocity of $v_s = 2000 \,\text{km} \, \text{s}^{-1}$.
The pulsar wind was injected at a speed of $v_p=0.99 \, \text{c}$ and a mass-loss rate scaled with $\eta = 0.1$ in \cref{eq: pulsar mass loss rate}.
\\
The simulation was performed on a Cartesian grid in corotating coordinates (as detailed in \cref{sec: orbital motion}).
The computational volume has the dimensions $[-0.4,0.4]\times[-0.2,0.6]\times[-0.2,0.2]\,$AU$^3$ and is homogeneously subdivided into $256\times 256\times 128$ spatial cells.
The binary orbits in the $x-y$ plane, with the stellar companion residing at the coordinate origin.
\\
The stellar and pulsar winds are injected as described in \cref{sec: wind injection} with an injection radius of four cell widths.
Initially, both injection volumes were placed in a vacuum.
We ran the simulation for $0.58\,$d giving the slow stellar wind time to populate the computational domain.

\subsection{Fluid structure}\label{sec: fluid results}
\begin{figure}
  \centering
  \includegraphics[width=\linewidth]{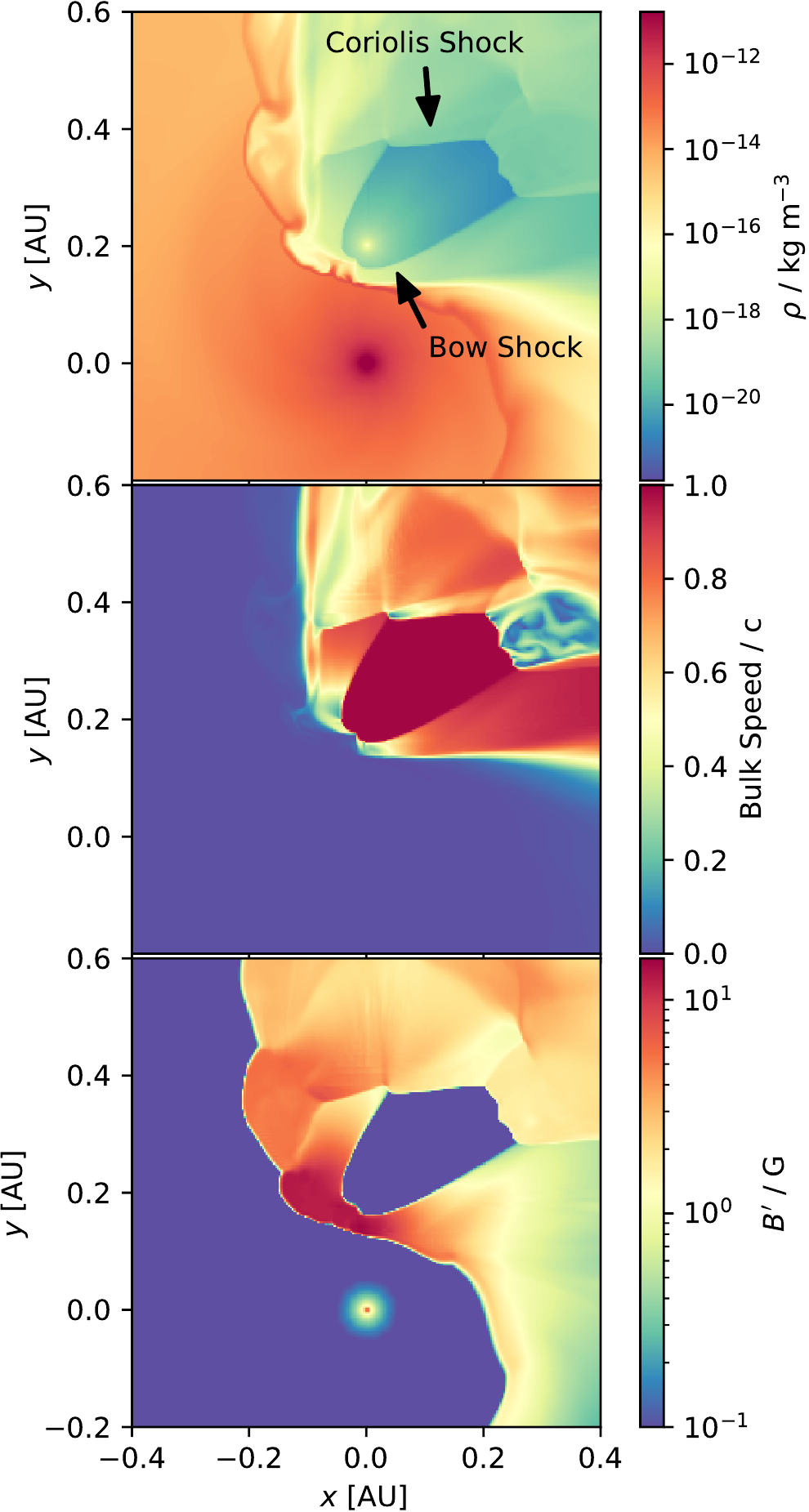}
  \caption{
    Resulting fluid structure for a generic binary system in the orbital plane.
    From top to bottom, we show the fluid mass density, bulk four-speed, and magnetic field strength in the fluid frame.
    An extended WCR is formed by interaction of the pulsar (blue) and stellar (orange) wind.
    The reference frame is corotating with counter-clockwise orbital motion.
    The location of the most relevant shocks is annotated.
  } \label{fig: toy fluid}
\end{figure}In \cref{fig: toy fluid} we show the resulting fluid mass, bulk four-speed, and magnetic field strength in the fluid frame, using the magnetic field model as described in \cref{sec: magnetic field model} with $\zeta_b = 0.5$.
The interaction of the stellar and pulsar wind gives rise to an extended WCR delimited by shocks on both the stellar and the pulsar side.
Through the Coriolis force in the rotating system, the WCR is highly asymmetric and bends in the direction opposite to the orbital motion.
This leads to the termination of the pulsar wind on the far side of the system, forming another shock at the Coriolis turnover (hereafter Coriolis shock).
The location of the Coriolis shock was noted previously and is consistent with previous studies \citep[see e.g.][]{Bosch-Ramon2015A&A...577A..89B, Dubus2015A&A...581A..27D}.
In the wings of the WCR, the shocked pulsar wind is reaccelerated to a Lorentz factor $\sim 3$, again in agreement with previous simulations by \citet{Bogovalov2008MNRAS.387...63B}.
\\
The strong velocity shear between the shocked winds at the contact discontinuity and the steep density gradient at the stellar wind shock leads to the development of fluid instabilities \citep[see also][]{Bosch-Ramon2012A&A...544A..59B}.
As a consequence, many secondary shocks are formed in the turbulent regions.

\subsection{Particle distribution}\label{sec: particle results}
To economise on computation time, we did not perform a particle transport simulation for the full time span used in \cref{sec: fluid results}.
Instead, we restarted such a simulation on a preconverged fluid solution and ran it for a shorter time span $t=0.029\,$d, allowing the accelerated electron-positron pairs to populate the system, which we also refer to as electrons for brevity.
\\
The simulation presented in this section was carried out using 50 logarithmic bins in energy ranging from $\gamma \in [200, 4 \times 10^8]$.
The particle injection was controlled as detailed in \cref{sec: particle injection} using $\zeta_\rho = 5 \times 10^{-4}$, $\zeta_n^\text{PL} = 5 \times 10^{-3}$, $\zeta_e^\text{PL} = 0.5$, $s = 2,$ and $\xi_\text{acc} = 1$, assuming the maximum acceleration timescale at the Bohm limit.
\\
\begin{figure}
  \centering
  \includegraphics[width=\linewidth]{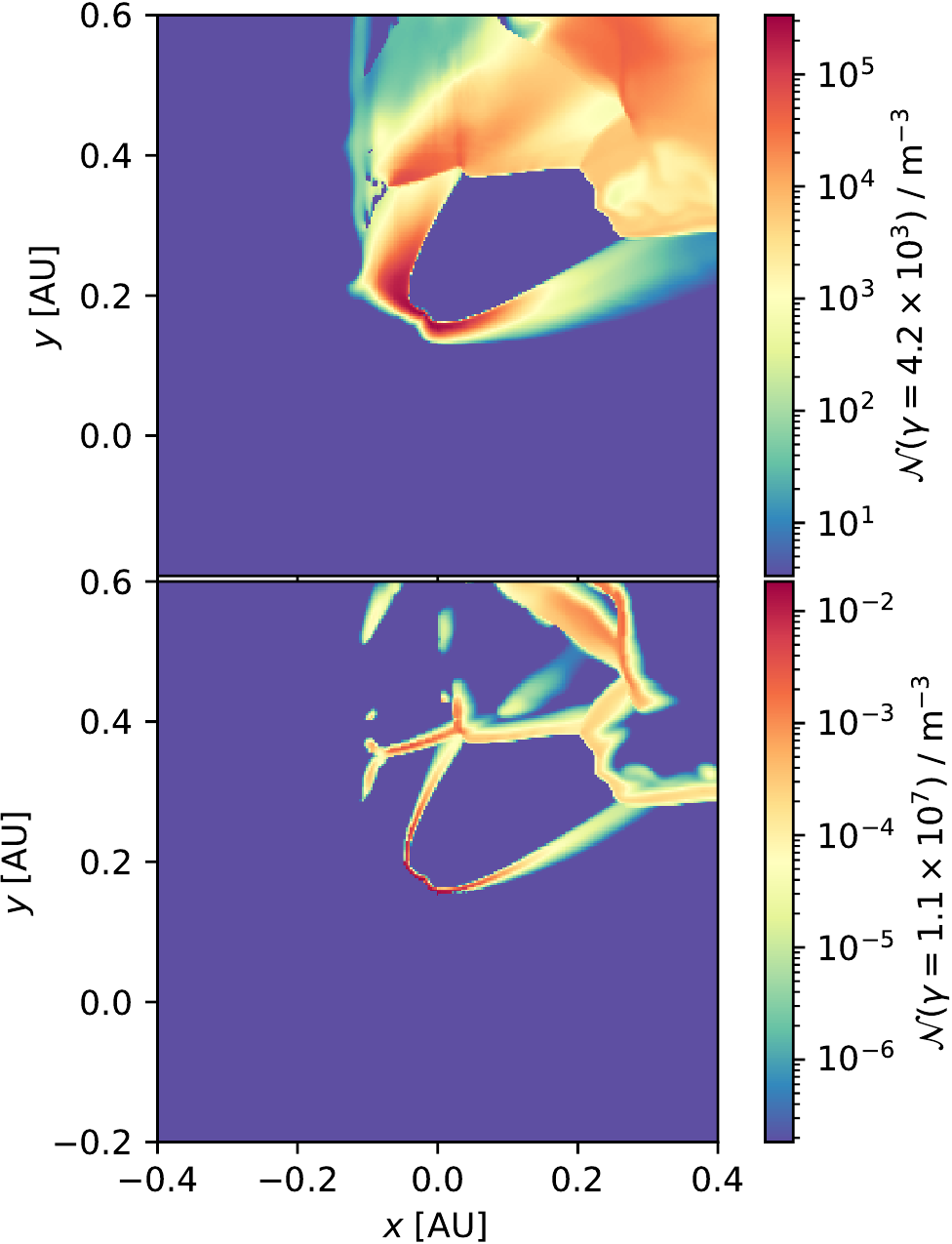}
  \caption{
    Differential particle number density for the energy bin centred at $\gamma = 4.2 \times 10^3$ (upper panel) and $\gamma = 1.1 \times 10^7$ (lower panel) in analogy to \cref{fig: toy fluid}.
    the lower energetic distribution is mainly populated by Maxwellian electrons, whereas the higher energetic distribution is dominated by power-law electrons.
    For clarity we show only five orders of magnitude below the highest value. The blue regions correspond to those with lower values.
  } \label{fig: toy particles}
\end{figure}\begin{figure}
  \centering
  \includegraphics[width=\linewidth]{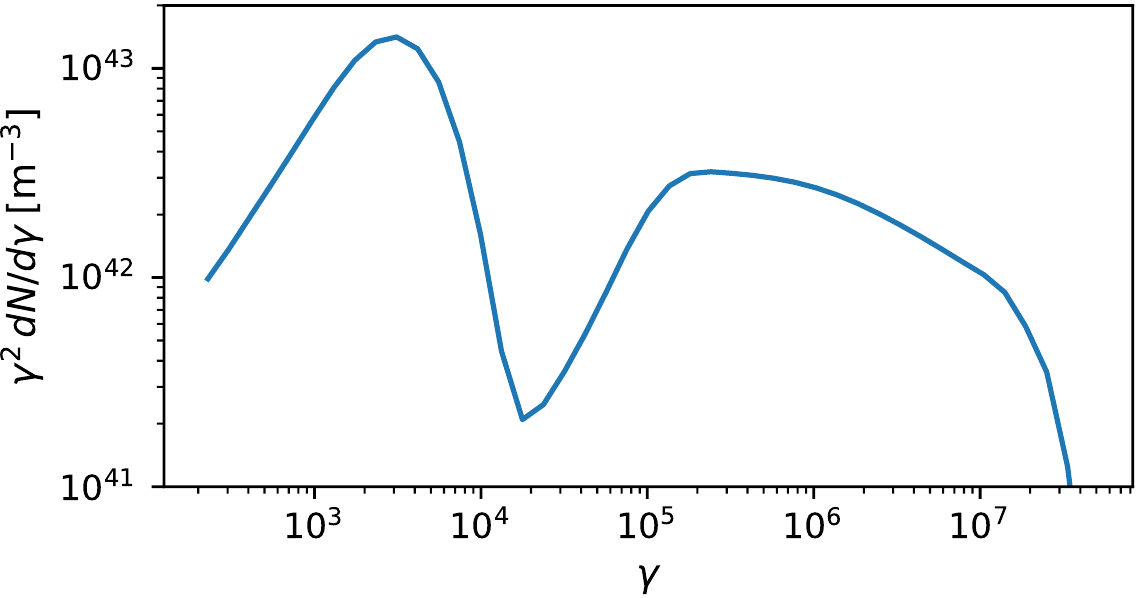}
  \caption{
    Spectral energy distribution of accelerated electrons integrated over the computational domain for different orbital phases.
  } \label{fig: toy particles spectrum}
\end{figure}In \cref{fig: toy particles} we present the resulting electron distributions at two energies.
The shock structure leaves a clear imprint on the particle distribution.
This is especially visible for higher energetic electrons, which remain much more confined to their acceleration site due to increased energy-loss rates, as compared to lower energetic electrons.
Next to the bow and Coriolis shocks, the secondary shocks arising in the turbulent flow are clearly visible, which are for the first time consistently accounted for in the particle transport.
\\
In \cref{fig: toy particles spectrum} we show the spectral energy distribution of electrons integrated over the computational volume.
The contributions of the Maxwellian and the power-law electrons can be easily distinguished, with Maxwellian electrons dominating the low-energy part ($\gamma \lesssim  2 \times 10^4$) and the non-thermal electrons the high-energy part.
For the chosen set of parameters, Maxwellian electrons were on average injected with $\gamma_t \sim 1.0 \times 10^3$ and non-thermal electrons within the limits of $\gamma_\text{min} \sim 1.1 \times 10^5$ and $\gamma_\text{max} = 2.4 \times 10^7$.
\\
The number and extent of particle acceleration sites identified in the turbulent Coriolis shock downstream region were determined by the chosen compression threshold in \cref{eq: injection criterion} (see also \cref{app: injection criterion}).
The particles accelerated in this region reach higher energies because the magnetic field is lower than in regions near the bow shock.
Choosing a more restrictive threshold, that is, one with higher compression, would thus decrease the total number of particles at higher energies in the simulation.

\subsection{Emission}
In this section, we present the predicted emission for the generic binary.
The involved computations were performed in a post-processing step as detailed in \cref{sec: gamma ray computation}, using the previously obtained particle distribution (see \cref{sec: particle results}) and fluid solution (see \cref{sec: fluid results}).
\\
We chose the observer to be located at a distance of $d_\text{obs} = 2.5\,$kpc along the $+y$ -axis of the system for the orbital phase $\phi=0$.
The phases $\phi=0$ and $\phi=0.5$ therefore correspond to inferior and superior conjunction, respectively.
As a result of the employed corotating coordinates, the observer rotates clockwise with respect to the simulation frame.
For the present purpose, the underlying particle and fluid solutions were kept constant for all orbital phases.
\\
To compute the inverse-Compton emission, we treated the stellar photon field as monochromatic, whereas for the $\gamma \gamma$ -absorption, a full blackbody spectrum was considered.
In both cases, the source was treated as an extended sphere with luminosity $L_\star = 2 \times 10^5\, \text{L}_\odot$, temperature $T_\star = 40000\,\text{K,}$ and resulting radius $R_\star = 9.3 \, \text{R}_\odot$.
The emission was computed for a system inclination of $i=30^\circ$ and 20 equidistant orbital phases.
\\
\begin{figure}
  \centering
  \includegraphics[width=\linewidth]{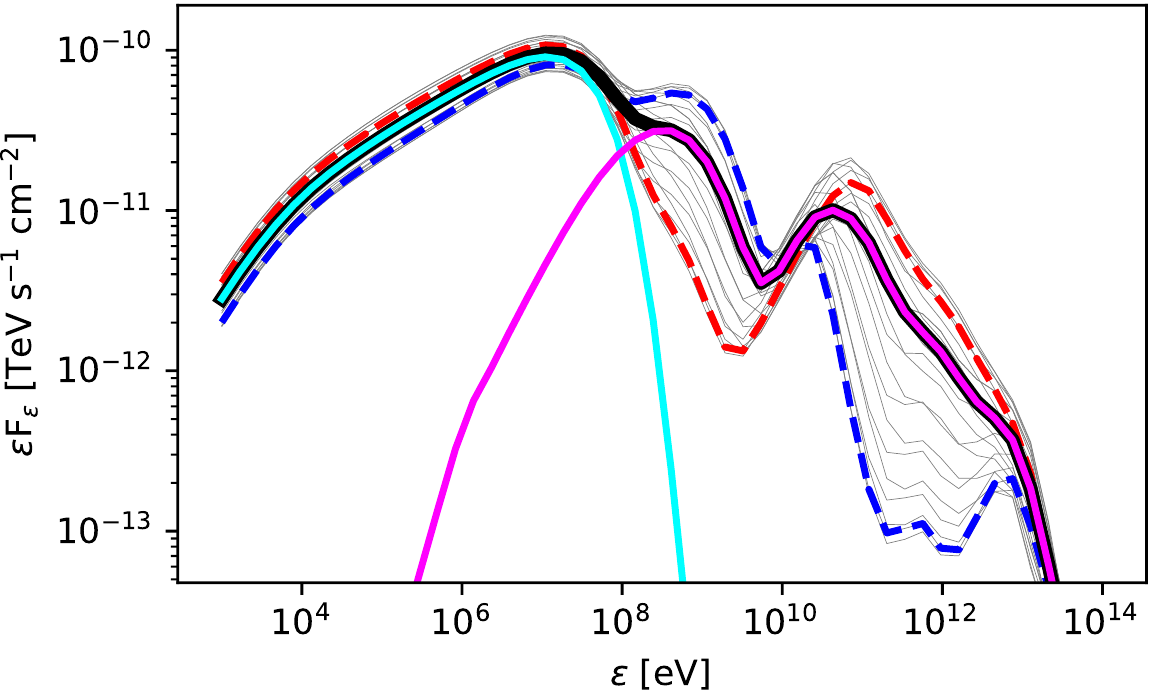}
  \caption{
    Spectral energy distribution of the emission predicted by our model for a generic binary with a system inclination of $30^\circ$.
    The results are shown for the sampled orbital phases (grey) and their average (black).
    The average spectral distribution is further split into the individual radiative processes: synchrotron (green) and inverse Compton attenuated by $\gamma \gamma$ -absorption (orange).
    Inferior (dashed red, $\phi=0$) and superior conjunction (dashed blue, $\phi=0.5$) emissions are highlighted.
  } \label{fig: toy spectrum}
\end{figure}In \cref{fig: toy spectrum} we show the resulting spectral energy distribution of emitted photons.
Ranging from X-rays up to LE gamma-rays, the spectrum is dominated by synchrotron emission from electrons at the power-law tail of the spectrum.
The same population of electrons is responsible for the inverse-Compton scattering of stellar UV photons to the VHE gamma-ray regime, which are heavily attenuated by $\gamma \gamma$-absorption in the $0.1-10\,$TeV range, depending on the orbital phase.
Again, the emission in the HE gamma-ray regime is produced through the inverse-Compton process, but by the low-energy Maxwellian electrons.
\\
In our model, the predicted spectra can be tuned most directly by varying the parameters that regulate the emerging electron distributions, that is, the unknown injection parameters and the magnetic field model.
Here, the acceleration efficiency $\xi_\text{acc}$ limits the maximum energy reached $\gamma_\text{max}$ (see \cref{eq: injection gamma max}), hence affecting the location of the LE and VHE cutoff.
The shape of the injected electron spectrum, and consequently, the main emission features, are determined by the electron specific energy density given by the ratio $\zeta_e^\text{PL,MW} / (\zeta_\rho \zeta_n^\text{PL,MW})$.
In the case of the Maxwellian electrons, this can be used to shift the inverse-Compton bump at HE in energy.
For the non-thermal power-law electrons, their specific energy density affects the low-energy cutoff of the spectrum.
The overall normalisation of the spectrum is given by the amount of energy deposited in the respective populations, determined by $\zeta_e^\text{PL,MW}$.
The index of the injected electrons translates into the spectral index of the X-ray synchrotron emission and affects the spectral index at VHE gamma-rays.
$\zeta_b$ is used to control the magnetic field strength in the simulation that affects the amplitude of the synchrotron emission, the VHE cutoff, and the evolution of the electron distribution itself with more subtle consequences for the emission spectra.
Because the compression threshold $\delta_\mathrm{sh}$ mostly affects the particle density at highest energies in the turbulent Coriolis shock downstream, this parameter can affect the synchrotron fluxes produced in the LE gamma-ray band, depending on the magnetic field in this region, and the VHE gamma-ray flux at superior conjunction.
\\
\begin{figure}
  \centering
  \includegraphics[width=\linewidth]{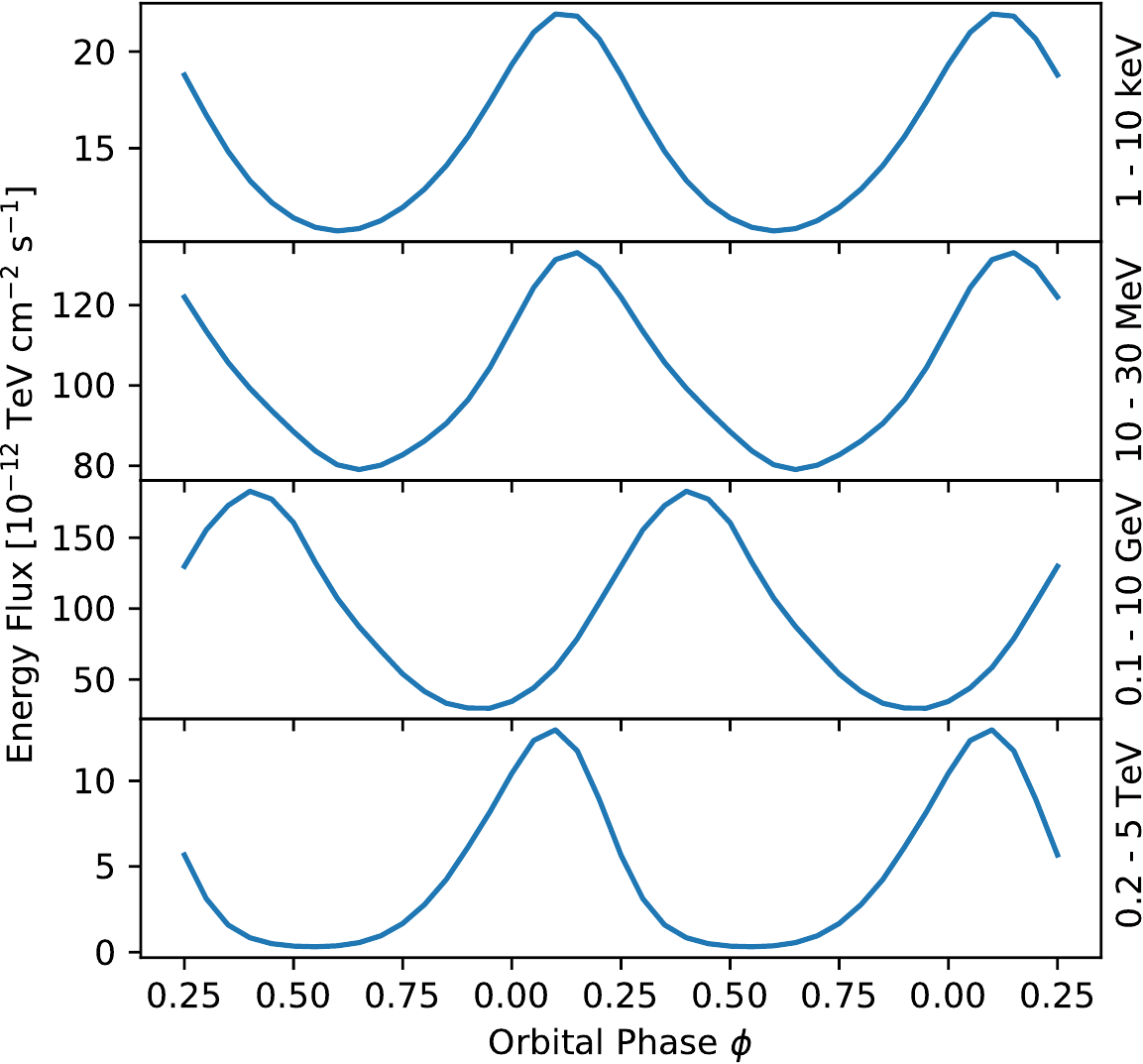}
  \caption{
    Emission light curves predicted by our model for a generic binary for different energy bands and a system inclination of $30^\circ$.
    Inferior and superior conjunction correspond to $\phi=0$ and  $\phi=0.5$, respectively.
    For better visualisation, we show the data for two full orbits.
  } \label{fig: toy lightcurve}
\end{figure}In \cref{fig: toy lightcurve} we show the predicted light curves for different energy bands, which exhibit a significant orbital modulation in all bands.
The X-ray and LE gamma-ray light curves show a clear correlation, which is expected because both bands are dominated by synchrotron emission generated by the same particle population.
Because we assumed no direction for the magnetic field, the synchrotron emissivity is isotropic in the fluid frame of our model.
The driver for the orbital modulation in these bands is accordingly solely given by relativistic boosting.
Because the WCR is bent by orbital motion, relativistic boosting is at its maximum at $\phi \sim 0.1$, while deboosting is most significant for $\phi \sim 0.6$.
\\
At HE gamma-rays, the anisotropic inverse-Compton scattering cross-section introduces a further line-of-sight-dependent modulation.
The scattering process is most efficient for large scattering angles, which are achieved at superior conjunction ($\phi = 0.5$).
This is reflected in the respective light curve, which reaches its maximum shortly before superior conjunction through the onset of relativistic deboosting, yielding an anti-correlation of the HE gamma-ray band and the previously discussed one.
\\
The VHE gamma-ray emission is again produced by inverse-Compton process, but the upscattered photons are attenuated by $\gamma\gamma$-pair creation with the stellar photon field.
This once more introduces a line-of-sight-dependent modulation that reaches its maximum attenuation at superior conjunction, where the produced VHE photons have to propagate through the entire stellar radiation field.
Consequently, the maximum photon flux can escape the system at inferior conjunction, with a slight shift to $\phi = 0.1$ due to relativistic boosting, leaving the VHE correlated with the X-ray to LE gamma-ray band and therefore anti-correlated to HE gamma-rays. \section{Summary and outlook}\label{sec: summary}
We presented a novel numerical model for gamma-ray binaries treating the transport of shock-accelerated electron-positron pairs dynamically coupled to the relativistic fluid dynamics of the stellar and pulsar wind interaction.
With this, it is possible for the first time to obtain self-consistent three-dimensional particle distributions and fluid solutions while consistently accounting for dynamic fluid instabilities, mixing, and the effects of orbital motion. 
The proposed emission model is purely leptonic, thus we inferred the expected emission through the synchrotron and inverse-Compton process acting on accelerated pairs from the pulsar wind obtained through the particle transport simulation.
Together with the simultaneously obtained fluid solution, it is therefore possible to consistently account for relativistic boosting and $\gamma \gamma$-absorption.
The model was implemented as a relativistic extension to the \cronos code \citep{Kissmann2018ApJS..236...53K}.
\\
The capabilities of the presented model were demonstrated on a generic binary system.
The wind interaction yields an extended asymmetric wind collision region bent by the effects of orbital motion, exhibiting a strong bow-like pulsar-wind shock and a Coriolis shock behind the pulsar.
The expressed features are in qualitative agreement with previous works \citep[see e.g.][]{Bosch-Ramon2015A&A...577A..89B}.
\\
The developing fluid instabilities give rise to the formation of secondary shocks.
Our model naturally includes the reacceleration of cooled particles at these shocks, which has not been done before in the case of gamma-ray binaries.
In our generic simulation, parts of the turbulence are damped by numerical diffusion due to our choices of spatial resolution.
More highly resolved simulations might hence yield a more turbulent fluid and consequently more secondary shocks.
\\
The emission predicted for the generic binary was extended over a broad range in energy reaching up to VHE gamma-rays.
Our results are consistent with the simulations performed by \citet{Dubus2015A&A...581A..27D}, showing a strong resemblance because the investigated models are similar.
\\
Our model predicts significant orbital modulation in every energy band.
The main driver for the modulation in the X-ray to LE gamma-ray synchrotron emission is given by relativistic boosting in the wings of the WCR.
In addition to the modulation introduced by the anisotropic inverse-Compton scattering producing the HE and VHE gamma-ray emission, the VHE band is further modulated by $\gamma \gamma$-absorption.
For the investigated general binary, the model naturally predicts a correlated X-ray to LE gamma-ray and VHE gamma-ray emission, while being anti-correlated to the HE band, as is seen in the observations of many gamma-ray binaries 
\citep[e.g. LS 5039][]{Chang2016MNRAS.463..495C}. 
\\
Because many of the observed gamma-ray binaries are known to host Be stars, a possible extension of the model foresees the inclusion of more complex, anisotropic stellar wind structures, such as adding an equatorial disc.
In the forthcoming second paper of this series, we will apply the presented model specifically to the well-studied LS 5039 system \citep{Huber2020-2}, for which no stellar disc has been found.
The wealth of available data means that the comparison of our model predictions with observations will provide valuable insights into gamma-ray binaries and a calibration of our model, pointing out potential aspects that might need more modelling effort in the future.
\\
This might, for example, consist of an inclusion of the pulsar and stellar magnetic fields in the dynamics of the wind interaction by extending the model to relativistic magnetohydrodynamics.
This might be used to obtain a self-consistent picture for the magnetic field strength and direction, which could further enable the incorporation of more sophisticated particle acceleration models in the future, for example including the obliquity of magnetised shocks \citep[e.g.][and references therein]{Vaidya2018ApJ...865..144V}.
\\
Another possible extension of this model could consist of easing some simplifications made in the particle transport model, for example by including spatial diffusion.
Next to the obvious changes in the spatial particle distribution, an additional effect is provided by the loss term accounting for non-inertial frame changes.
This effect has not been studied in the context of gamma-ray binaries.
\\
In future, our model can further help to shed light on dynamical phenomena that can now be addressed with a higher level of sophistication.
This could involve investigations of the role of turbulence and the variability it introduces in the radiative output of gamma-ray binaries.
Particularly, it remains to be investigated in which way the emission from these systems may vary from one orbit to the next.
In a broader context, this might also yield implications for and connect to transient phenomena, such as flaring events in otherwise regular gamma-ray binary emissions \citep[e.g. PSR B1259-63][]{Johnson2018ApJ...863...27J}. 
\begin{acknowledgements}
   The computational results presented in this paper have been achieved (in parts) using the research infrastructure of the Institute for Astro- and Particle Physics at the University of Innsbruck, the LEO HPC infrastructure of the University of Innsbruck, the MACH2 Interuniversity Shared Memory Supercomputer and PRACE resources.
   We acknowledge PRACE for awarding us access to Joliot-Curie at GENCI@CEA, France.
   This research made use of Cronos \citep{Kissmann2018ApJS..236...53K}; GNU Scientific Library (GSL) \citep{galassi2018scientific}; matplotlib, a Python library for publication quality graphics \citep{Hunter2007}; and NumPy \citep{van2011numpy}.
   We would like to thank the unknown referee/reviewer for the thoughtful comments and efforts towards improving our manuscript.
\end{acknowledgements}

\bibliographystyle{aa}
\bibliography{references}

\begin{appendix}
   
\section{Corotating frame} \label{app: corotating frame}
We considered a reference frame that corotates with the mean angular velocity of the binary $\Omega$ in the $x-y$ plane.
The corotation velocity is in this case given by the orbital period of the system as $\Omega = 2 \pi / P_\text{orb}$.
The set of new coordinates, indicated by a bar, can be readily expressed as
\begin{align}
  \bar{t} & = t \label{eq: corot trafo 1}                                     \\
  \bar{x} & = x \cos \left( \Omega t \right) + y \sin \left( \Omega t \right) \\
  \bar{y} & = y \cos \left( \Omega t \right) - x \sin \left( \Omega t \right) \\
  \bar{z} & = z. \label{eq: corot trafo 4}
\end{align}
Transforming the flat Cartesian Minkowski metric into the corotating system yields
\begin{align}
  g_{\bar{\mu} \bar{\nu}}=
    {\arraycolsep=0.4\arraycolsep\ensuremath{
        \begin{pmatrix}
          - 1 + \Omega^2 \left( \bar{x}^2 + \bar{y}^2 \right) & - \Omega \bar{y} & \Omega \bar{x} & 0 \\
          -\Omega \bar{y}                                     & 1                & 0              & 0 \\
          \Omega \bar{x}                                      & 0                & 1              & 0 \\
          0                                                   & 0                & 0              & 1
        \end{pmatrix}
      }}\label{eq: corot metric}
   \\
  g^{\bar{\mu} \bar{\nu}} =
    {\arraycolsep=0.4\arraycolsep\ensuremath{
        \begin{pmatrix}
          -1               & - \Omega \bar{y}         & \Omega \bar{x}           & 0 \\
          - \Omega \bar{y} & 1 - \Omega^2 \bar{y}^2   & \Omega^2 \bar{x} \bar{y} & 0 \\
          \Omega \bar{x}   & \Omega^2 \bar{x} \bar{y} & 1 - \Omega^2 \bar{x}^2   & 0 \\
          0                & 0                        & 0                        & 1
        \end{pmatrix}.
      }} \label{eq: corot metric inverse}
\end{align}

Generic metric tensors can be accounted for in the broader framework of general relativistic hydrodynamics.
Here, the equations for energy and momentum conservation are given by $\nabla_\mu T^{\mu \nu} =0$ with the energy-momentum tensor $T^{\mu \nu} = \rho h u^\mu u^\nu + p g^{\mu \nu}$.
Because of the non-vanishing, off-diagonal terms in the metric, the SRHD schemes can no longer be applied without modification.
\\
In a 3+1 decomposition \citep{MisnerGravitation1973}, the line element is expressed as $\td s^2 = -(\alpha^2 - \beta_i \beta^i) \td t^2 + 2 \beta_i \td x^i \td t + \gamma_{ij} \td x^i \td x^j$, with $\bar{\alpha}$ the lapse, $\bar{\beta}^i$ the shift, and $\bar{\gamma}_{ij}$ the spatial metric.
These gauge functions can be directly obtained from the metric tensor \cref{eq: corot metric} and \cref{eq: corot metric inverse}.
For the presented corotating system, they are given by
\begin{align}
  \bar{\alpha} = 1                                                                           \\
  \bar{\beta}^i = \bar{\beta}_i = \left( - \Omega \bar{y}, \, \Omega \bar{x}, \, 0 \right)_i \\
  \bar{\gamma}_{ij} = \mathds{1.}
\end{align}
In our approach, we relied on the so-called \textit{\textup{3+1 Valencia}} formulation of general relativistic hydrodynamics \citep{Banyuls1997ApJ...476..221B}, and chose the conserved variables 
\begin{equation}
  \vU = 
  \begin{pmatrix}
    D \\ E \\ m_j
  \end{pmatrix} =
  \begin{pmatrix}
    \rho W \\ D h W - p \\ D h W w_i
  \end{pmatrix},
  \label{eq: valencia conserved}
\end{equation}
consisting of the rest-mass density $D$, the momentum density in the $j$-direction $m_j$, and the total energy density $E$, measured by a family of Eulerian observers.
In their respective frames, the fluid three-velocity is given by $w^i$ , which is related to the four-velocity in the rotating system by $w^i = (u^{\bar{i}}/u^{\bar{t}} + \beta^{\bar{i}}) / \alpha$.
The corresponding Lorentz factor is given by $W = \left(1 - w_i w^i \right)^{-1/2}$ with $w_i=\gamma_{ij} w^j$.

The chosen variables closely resemble those in SRHD \cref{eq: srhd conserved} up to the usage of covariant momentum densities instead of contravariant ones, which are equivalent and interchangeable in our case because $\gamma_{ij} = \mathds{1}$.
This allows an effective reuse of the variable inversion scheme developed for the special relativistic case that is already implemented in \cronos.
\\
The evolution equations take the generic form 
\begin{equation}
  \partial_t (\sqrt{\gamma} \vU) + \partial_i (\sqrt{\gamma} \vF^i) = \sqrt{\gamma}\mathbf{S} \label{eq: valencia system}
,\end{equation}
with $\gamma = \det (\gamma_{ij})$ and the fluxes
\begin{equation}
  \vF^i = 
  \begin{pmatrix}
    \alpha w^i D  - \beta^i D\\ \alpha m^i - \beta^i E \\ \alpha(m_j w^i + p \delta^i_j) - \beta^i m_j \label{eq: valencia fluxes}
  \end{pmatrix}.
\end{equation}
The system of equations again closely resemble those in special relativity \cref{eq: srhd conservation equation} up to the additional fluxes introduced by $\beta^i \neq 0$.
We treated the equations using the scheme presented by \cite{Pons1998A&A...339..638P}, describing how special relativistic Riemann solvers can be used with general metrics.
The required transformations between the Eulerian and the local inertial frames in our case take the especially simple form
\begin{equation}
  {M^{\hat{i}}}_i \vert_{ \Sigma_{\bar{x}}} = \mathds{1}, \quad 
  {M^{\hat{i}}}_i \vert_{ \Sigma_{\bar{y}}} = \mathds{1}, \quad 
  {M^{\hat{i}}}_i \vert_{ \Sigma_{\bar{z}}} = \mathds{1}.
\end{equation}
The source terms in \eqref{eq: valencia system} are given by \citep{Banyuls1997ApJ...476..221B}
\begin{equation}
  \mathbf{S} = 
  \begin{pmatrix}
    0 \\
    \alpha^2 \left( T^{\mu t} \partial_\mu \ln \alpha - \Gamma^t_{\mu \lambda} T^{\mu\lambda} \right) \\
    \alpha T^{\mu \nu} \left( \partial_\mu g_{\nu j} - \Gamma^\lambda_{\nu \mu} g_{\lambda j} \right) 
  \end{pmatrix},
  \label{eq: valencia sources}
\end{equation}
and reduce to
\begin{equation}
  {\bf S} = \left( 0, \, 0, \, \Omega \, m_{\bar{y}}, \, -\Omega \, m_{\bar{x}}, \, 0 \right)^\top
\end{equation}
in our case.
This result may seem counter-intuitive at first glance because the terms formally do not correspond to their Newtonian analogue for the centrifugal and Coriolis force in a rotating system.
However, we recall that the conserved quantities are measured by the Eulerian observers instead of in the rotating frame.
\\
Finally, our approach is equivalent to others such as \citet{Papadopoulos2000PhRvD..61b4015P}, who chose $T^{0 \nu}$ as conserved variables and evolved the quantities directly in the rotating frame, which recovers the well-known terms for the fictitious forces.
This formulation, however, is less efficient in practice because the involved fluxes are more complex, and the whole metric has to be considered for contractions, for example in the computation of Lorentz factors, instead of just the spatial metric in the \textit{\textup{3+1 Valencia}} formulation, which is unity in our case.
Furthermore, the chosen conserved variables formally differ from \cref{eq: srhd conserved}, prohibiting the reuse of already developed SRHD variable inversion schemes \citep[e.g.][]{Mignone2007MNRAS.378.1118M}.
\\
Ultimately, we are interested in quantities measured in the stationary laboratory frame to compute the radiative output of the system.
This amounts to a Lorentz transformation of the corotating velocity field $u^{\bar{\mu}}$, which is obtained from the velocities measured by the Eulerian observers to the stationary $u^{\mu}$.
For our case, we find after some algebra that the Eulerian velocities directly correspond to those in the stationary system up to a rotation in space.
This rotation is accounted for in the computation of relativistic boosting described in \cref{sec: gamma ray computation}.

\section{Radiation fields}\label{sec: radiation fields}
Depending on the radiation field at hand, the integrals in the computation of inverse-Compton emission or $\gamma \gamma$ absorption can be simplified.
For example, for monochromatic seed photon fields, the energy integration reduces to a trivial substitution.
To be able to map these properties over to our implementation, we employed a factorisation of the radiation field.
In general, a radiation field can be described as the number of photons $n$ per unit volume, unit energy, and unit solid angle, depending on the location $\mathbf{x}$, the photon energy $\epsilon$, and the direction $\mathbf{\Omega}$.
For our purposes, however, the spectral shape is direction independent, which allows the following factorisation of the photon density
\begin{equation}
  n(\mathbf{x}, \epsilon, \mathbf{\Omega}) = \frac{\td N}{\td V \td \epsilon \td \Omega} =
  \rfn(\mathbf{x})
  \cdot \rfe(\mathbf{x}, \epsilon)
  \cdot \rfd(\mathbf{x}, \mathbf{\Omega})
,\end{equation}
with a factor $\rfn$ describing the total photon density, a factor $\rfe$ describing the energy dependence, and a factor $\rfd$ describing the directional profile,
\begin{equation}
  \rfn(\mathbf{x}) := \frac{\td N}{\td V}
\end{equation}
\begin{equation}
  \rfe(\mathbf{x}, \epsilon) := \left( \frac{\td N}{\td V} \right)^{-1} \frac{\td N}{\td V \td \epsilon}
\end{equation}
\begin{equation}
  \rfd(\mathbf{x}, \mathbf{\Omega}) := \left( \frac{\td N}{\td V} \right)^{-1} \frac{\td N}{\td V \td \Omega}
.\end{equation}
We note that $\rfe$ and $\rfd$ are normalised with respect to $\epsilon$ and $\mathbf{\Omega}$, respectively.
\\
When we consider spherically expanding photons injected at a location $\mathbf{x}_\star$ with a given rate $\dot{N}_\star$ , the photon number density is given by
\begin{equation}
  \rfn^\text{spherical}(\mathbf{x}) = \frac{1}{r^2} \frac{\dot{N}_\star}{4 \pi c}
  \label{eq: factor spherical}
,\end{equation}
where $r = \norm{\mathbf{x} - \mathbf{x}_\star}$.
\\
A mono-directional or beamed radiation field in a given direction $\mathbf{\Omega}_\text{beam}$ is in general described by
\begin{equation}
  \rfd^\text{beamed}(\mathbf{x},\mathbf{\Omega}) = \delta \left( \mathbf{\Omega} - \mathbf{\Omega}_\text{beam} \right)
  \label{eq: factor beamed}
.\end{equation}
The radiation field of a point-like source can therefore be viewed as a special case of a field beamed in the radial direction $\mathbf{\Omega}_\text{beam} = \mathbf{e}_r = \mathbf{r}/r$.
An extended spherical source is approximated as a disc with constant brightness and angular extent limited by the source radius $R$,
\begin{equation}
  \rfd^\text{extended}(\mathbf{x},\mathbf{\Omega}) =
  \begin{cases}
    \left( 2\pi (1 - \mu_{0,\text{min}}) \right)^{-1} & \mu_0 > \mu_{0,\text{min}} \\
    0                                                 & \text{else}
  \end{cases}
  \label{eq: factor extended}
,\end{equation}
where $\mu_0$ is the cosine of the angle enclosed between $\mathbf{\Omega}$ and $\mathbf{e}_r$ and $\mu_{0,\text{min}} =  \left(1 - R^2/r^2 \right)^{1/2}$.
\\
The radiation field of the companion star can be approximated as a blackbody spectrum with temperature $T,$
\begin{equation}
  \rfe^\text{blackbody}(\mathbf{x},\epsilon) = \frac{1}{2 \zeta(3) k_b T}\frac{(\epsilon / k_b T)^2}{\exp(\epsilon / k_b T) - 1}
  \label{eq: factor black body}
.\end{equation}
This can be further approximated as a monochromatic spectrum with the average photon energy $\bar{\epsilon} = k_B T \frac{\pi^4}{30 \, \zeta(3)}$,
\begin{equation}
  \rfe^\text{mono}(\mathbf{x},\epsilon) = \delta \left( \epsilon - \bar{\epsilon}\right)
  \label{eq: factor monochromatic}
.\end{equation}
\\
In our simulation, the radiation field of the star was parametrised by its luminosity $L_\star$ and its temperature $T_\star$.
The extent of the source is therefore given by $R_\star^2 = L_\star / ( \frac{4\pi}{3} \sigma T^4_\star)$ and its photon injection rate by $\dot{N}_\star = L_\star / \bar{\epsilon}_\star$. 
\section{Comparison to a common shock-identification criterion}\label{app: injection criterion}
In the literature, particle acceleration is thought to occur at strong shocks, which are usually identified by a diverging fluid flow and a jump in pressure \citep[see e.g.][]{Vaidya2018ApJ...865..144V}.
As described in \cref{sec: particle injection}, we employed a less restrictive criterion on the fluid compression alone because a jump in pressure might not yet have developed for a given instance in time.
\\
For the generic binary system considered in \cref{sec: generic binary}, we found both sets of criteria to identify the same main shock features, that is, the bow and Coriolis shock, with reflected shocks at their encounter and secondary shocks.
In comparison to our criterion, the usual approach identifies fewer and less extended acceleration sites in the turbulent Coriolis shock downstream region because of the additional pressure constraint.
This introduces a slight decrease in the emission at all wavelengths through the reduced particle injection, which could be compensated for by a different set of injection parameters.
Because the magnetic field in the Coriolis shock downstream region is lower, the effect becomes more relevant for particles at higher energies, leading to a decrease in the synchrotron emission in the LE gamma-ray band.
The temporal behaviour remains unaffected in all bands.
For a compression threshold of $\delta_\mathrm{sh} \sim -50 \, c / \mathrm{AU}$, we found that our criterion identified the same acceleration sites as the classical approach. We treated the threshold as a free parameter of the simulation, which might be constrained for a given system compared with observations. \end{appendix}

\end{document}